\documentclass[prl,twocolumn,floatfix,a4paper,superscriptaddress]{revtex4}
\usepackage{bm,color,graphicx,amsmath,txfonts}

\newcommand{\VV}{{\cal V}}

\begin{document}

\title{Magnon-photon-phonon entanglement in cavity magnomechanics}

\author{Jie Li}
\affiliation{Department of Physics, Zhejiang University, Hangzhou 310027, China}
\affiliation{Institute for Quantum Science and Engineering and Department of Biological and Agricultural Engineering, Texas A{\rm \&}M University, College Station, Texas 77843, USA}
\author{Shi-Yao Zhu}
\affiliation{Department of Physics, Zhejiang University, Hangzhou 310027, China}
\author{G. S. Agarwal}
\affiliation{Institute for Quantum Science and Engineering and Department of Biological and Agricultural Engineering, Texas A{\rm \&}M University, College Station, Texas 77843, USA}
\affiliation{Department of Physics and Astronomy, Texas A{\rm \&}M University, College Station, Texas 77843, USA}

\begin{abstract}
We show how to generate tripartite entanglement in a cavity magnomechanical system which consists of magnons, cavity microwave photons, and phonons. The magnons are embodied by a collective motion of a large number of spins in a macroscopic ferrimagnet, and are driven directly by an electromagnetic field. The cavity photons and magnons are coupled via magnetic dipole interaction, and the magnons and phonons are coupled via magnetostrictive (radiation pressure-like) interaction. We show optimal parameter regimes for achieving the tripartite entanglement where magnons, cavity photons, and phonons are entangled with each other, and we further prove that the steady state of the system is a genuinely tripartite entangled state. The entanglement is robust against temperature. Our results indicate that cavity magnomechanical systems could provide a promising platform for the study of macroscopic quantum phenomena.  
\end{abstract}

\date{\today}
\maketitle

In recent years ferrimagnetic systems, especially the yttrium iron garnet (YIG) sphere, have attracted considerable interest from the perspective of cavity quantum electrodynamics (QED). It is found that the Kittel mode~\cite{Kittel} in the YIG sphere can realize strong coupling with the microwave photons in a high-quality cavity, leading to cavity polaritons~\cite{Strong1,Strong2,Strong3,Strong4,Strong5} and the vacuum Rabi splitting. Thus many ideas originally developed in cavity QED can be applied to magnon cavity QED~\cite{Haroche,GA84,Hinds,Yao,Nori}. Other interesting developments in the context of magnon cavity QED are, e.g., the observation of bistability~\cite{You18} and the coupling of a single superconducting qubit to the Kittel mode~\cite{Science}. Clearly, magnon systems provide us with a new platform for studying unique effects of strong-coupling QED. This is very similar to other platforms provided by superconducting qubits~\cite{Wallraff}, semiconductor qubits~\cite{semicon}, and double quantum dots~\cite{2dots}.

The developments in cavity QED resulted in the birth of the new field of cavity optomechanics, where mechanical elements are coupled to the cavity via radiation pressure~\cite{OMRMP}. The field of cavity optomechanics is now being studied with many different systems such as superconducting elements~\cite{supercond1}. Recently, significant progress has been reported on the study of quantum effects, e.g., the quantum entanglement between mechanics and a cavity field~\cite{enOM}, as well as between two massive mechanical oscillators~\cite{enMM1,enMM2} have been observed. In the light of these advances, it is natural to investigate the utility of magnon systems in cavity optomechanics and their quantum characteristics. We note that the first realization of the magnon-photon-phonon interaction has been reported~\cite{Tang16}, where photons are coupled to magnons as in magnon QED and in addition magnons get coupled to phonons. The consequences of the magnon-phonon coupling are observed in the cavity output, but this study is at the mean field level, i.e., all quantum fluctuations are ignored.

Here we present a full quantum theory of the magnon-photon-phonon system. We show that it is possible to observe quantum effects, e.g., entanglement, between magnons, cavity photons, and phonons. Specifically, we show that, based on experimentally reachable parameters, not only all bipartite entanglements but also {\it genuine} tripartite entanglement could be generated in the magnon-photon-phonon system. All entanglements are robust against environmental temperature. The entanglement arises from the magnon-phonon coupling, without which it vanishes. We model the system by using the standard Langevin formalism, solve the linearized dynamics and quantify the entanglement in the stationary state. Finally, we analyze the validity of our model and show how to measure the generated entanglement.

\begin{figure}[t]
\hskip-0.08cm\includegraphics[width=\linewidth]{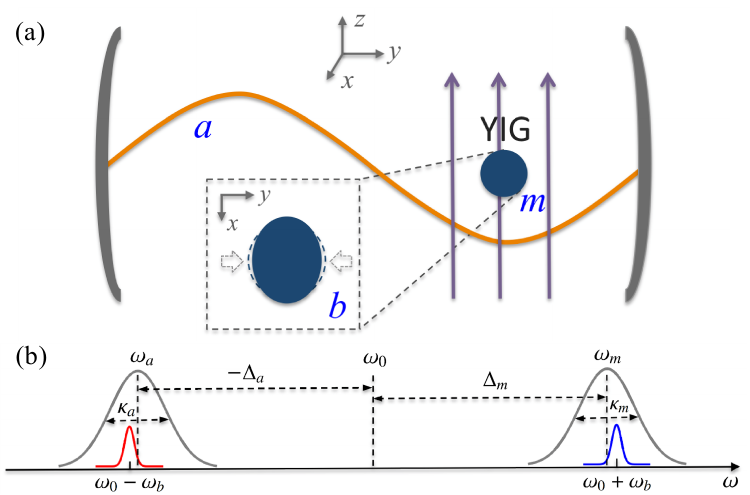} 
\caption{(a) Sketch of the system. A YIG sphere is placed inside a microwave cavity near the maximum magnetic field of the cavity mode, and simultaneously in an uniform bias magnetic field, which establish the magnon-photon coupling. The magnon mode is directly driven by a microwave source (not shown) to enhance the magnomechanical coupling. The bias magnetic field ($z$ direction), the drive magnetic field ($y$ direction) and the magnetic field ($x$ direction) of the cavity mode are mutually perpendicular at the site of the YIG sphere. (b) Frequencies and linewidths of the system. The magnon mode with frequency $\omega_m$ and bandwidth $\kappa_m$ is driven by a microwave field at frequency $\omega_0$ and the mechanical motion at frequency $\omega_b$ scatters photons onto the two sidebands at $\omega_0 \pm \omega_b$. If the magnon mode is resonant with the blue (anti-Stokes) sideband and the cavity with frequency $\omega_a$ and bandwidth $\kappa_a$ is resonant with the red (Stokes) sideband, the system exhibits genuine magnon-photon-phonon entanglement.}
\label{fig1}
\end{figure}

We consider a hybrid cavity magnomechanical system~\cite{Tang16} which consists of cavity microwave photons, magnons, and phonons, as shown in Fig.~\ref{fig1} (a). The magnons are embodied by a collective motion of a large number of spins in a ferrimagnet, e.g., a YIG sphere (a 250-$\mu$m-diameter sphere is used in Ref.~\cite{Tang16}). The magnetic dipole interaction mediates the coupling between magnons and cavity photons. The magnons couple to phonons via magnetostrictive interaction. Specifically, the varying magnetization induced by the magnon excitation inside the YIG sphere leads to the deformation of its geometry structure, which forms vibrational modes (phonons) of the sphere, and vice versa~\cite{Kittel2}. We consider the size of the sphere is much smaller than the microwave wavelength, such that the effect of radiation pressure is negligible. The Hamiltonian of the system reads
\begin{equation}
\begin{split}
H/\hbar &= \omega_a a^{\dag} a + \omega_m m^{\dag} m + \frac{\omega_b}{2} (q^2 + p^2) + g_{mb} m^{\dag} m q   \\
&+ g_{ma} (a + a^{\dag}) (m + m^{\dag}) + i \Omega (m^{\dag} e^{-i \omega_0 t}  - m e^{i \omega_0 t} ),
\end{split}
\end{equation}
where $a$ ($a^{\dag}$) and $m$ ($m^{\dag}$) ($[O, O^{\dag}]\,{=}\,1$, $O\,{=}\,a,m$) are the annihilation (creation) operator of the cavity and magnon modes, respectively, $q$ and $p$ ($[q, p]\,{=}\,i$) are the dimensionless position and momentum quadratures of the mechanical mode, and $ \omega_a$, $ \omega_m$, and $ \omega_b$ are the resonance frequency of the cavity, magnon and mechanical modes, respectively. The magnon frequency is determined by the external bias magnetic field $H$ and the gyromagnetic ratio $\gamma$, i.e., $\omega_m=\gamma H$. The magnon-microwave coupling rate $g_{ma}$ can be larger than the dissipation rates of the cavity and magnon modes, $\kappa_a$ and $\kappa_m$, entering into the strong coupling regime, $g_{ma} > \kappa_{a}, \kappa_{m}$~\cite{Strong1,Strong2,Strong3,Strong4,Strong5}. The single-magnon magnomechanical coupling rate $g_{mb}$ is typically small, but the magnomechanical interaction can be enhanced by driving the magnon mode with a strong microwave field (directly driving the YIG sphere with a microwave source has been adopted in Refs.~\cite{You18,You16}). The Rabi frequency $\Omega =\frac{\sqrt{5}}{4} \gamma \! \sqrt{N} B_0$~\cite{SM} denotes the coupling strength of the drive magnetic field (with amplitude $B_0$ and frequency $\omega_0$) with the magnon mode, where $\gamma/2\pi= 28$ GHz/T, and the total number of spins $N=\rho V$ with $V$ the volume of the sphere and $\rho=4.22 \times 10^{27}$ m$^{-3}$ the spin density of the YIG. Note that $\Omega$ is derived under the assumption of the low-lying excitations, $\langle m^{\dag} m \rangle \ll 2Ns$, where $s=\frac{5}{2}$ is the spin number of the ground state Fe$^{3+}$ ion in YIG.

In the frame rotating at the drive frequency $\omega_0$ and applying the rotating-wave approximation, $g_{ma} (a + a^{\dag}) (m + m^{\dag}) \to g_{ma} (a m^{\dag} + a^{\dag} m)$ (valid when $\omega_a, \omega_m \gg g_{ma}, \kappa_{a}, \kappa_{m}$, which is easily satisfied~\cite{Tang16}), the quantum Langevin equations (QLEs) describing the system are given by
\begin{equation}
\begin{split}
\dot{a}&= - (i \Delta_a + \kappa_a) a - i g_{ma} m + \sqrt{2 \kappa_a} a^{\rm in},  \\
\dot{m}&= - (i \Delta_m + \kappa_m) m - i g_{ma} a - i g_{mb} m q + \Omega + \sqrt{2 \kappa_m} m^{\rm in},  \\
\dot{q}&= \omega_b p,   \\
\dot{p}&= - \omega_b q - \gamma_b p - g_{mb} m^{\dag}m + \xi, 
\end{split}
\end{equation}
where $\Delta_{a}=\omega_{a}-\omega_0$, $\Delta_{m}=\omega_{m}-\omega_0$, $\gamma_b$ is the mechanical damping rate, and $a^{\rm in}$, $m^{\rm in}$ and $\xi$ are input noise operators for the cavity, magnon and mechanical modes, respectively, which are zero mean and characterized by the following correlation functions~\cite{Zoller}: $\langle a^{\rm in}(t) \, a^{\rm in \dag}(t')\rangle = [N_a(\omega_a){+}1] \,\delta(t{-}t')$, $\langle a^{\rm in \dag}(t) \, a^{\rm in}(t')\rangle \,\,{=}\,\, N_a(\omega_a) \, \delta(t{-}t')$, and $\langle m^{\rm in}(t) \, m^{\rm in \dag}(t')\rangle = [N_m(\omega_m)+1] \, \delta(t{-}t')$, $\langle m^{\rm in \dag}(t) \, m^{\rm in}(t')\rangle = N_m(\omega_m)\, \delta(t{-}t')$, and $\langle \xi(t)\xi(t')\,{+}\,\xi(t') \xi(t) \rangle/2 \,\, {\simeq} \,\, \gamma_b [2 N_b(\omega_b) {+}1] \delta(t{-}t')$, where a Markovian approximation has been made, which is valid for a large mechanical quality factor ${\cal Q} = \omega_b/\gamma_b \,\, {\gg}\, 1$~\cite{Markov} (a prerequisite for seeing quantum effects like entanglement), and $N_j(\omega_j){=}\big[ {\rm exp}\big( \frac{\hbar \omega_j}{k_B T} \big) {-}1 \big]^{-1} $ $(j{=}a,m,b)$ are the equilibrium mean thermal photon, magnon, and phonon number, respectively.

We assume that the magnon mode is strongly driven, leading to a large amplitude $|\langle m \rangle| \gg 1$ at the steady state, and due to the cavity-magnon beamsplitter interaction, the cavity field also has a large amplitude $|\langle a \rangle| \gg 1$. This allows us to linearize the dynamics of the system around the steady-state values by writing any operator as $O=\langle O \rangle +\delta O$ ($O\, {=}\, a,m,q,p$) and neglecting second order fluctuation terms. The linearized QLEs describing the quadrature fluctuations $(\delta X, \delta Y, \delta x, \delta y, \delta q, \delta p)$, with $\delta X=(\delta a + \delta a^{\dag})/\sqrt{2}$, $\delta Y=i(\delta a^{\dag} - \delta a)/\sqrt{2}$, $\delta x=(\delta m + \delta m^{\dag})/\sqrt{2}$, and $\delta y=i(\delta m^{\dag} - \delta m)/\sqrt{2}$, can be written as
\begin{equation}
\dot{u} (t) = A u(t) + n(t) ,
\end{equation}
where $u(t)=\big[\delta X (t), \delta Y (t), \delta x (t), \delta y (t), \delta q (t), \delta p (t) \big]^T$, $n (t) = \big[ \!\sqrt{2\kappa_a} X^{\rm in} (t), \sqrt{2\kappa_a} Y^{\rm in} (t), \sqrt{2\kappa_m} x^{\rm in} (t), \sqrt{2\kappa_m} y^{\rm in} (t), 0, \xi (t) \big]^T$ is the vector of input noises, and the drift matrix $A$ is given by  
\begin{equation}\label{AAA}
A =
\begin{pmatrix}
-\kappa_a  &  \Delta_a  &  0 &  g_{ma}  &  0  &  0   \\
-\Delta_a  & -\kappa_a  & -g_{ma}  & 0  &  0  &  0   \\
0 & g_{ma}  & -\kappa_m  & \tilde{ \Delta}_m &  -G_{mb}  &  0 \\
-g_{ma}  & 0 & -\tilde{ \Delta}_m & -\kappa_m &  0  &  0 \\
0 &  0  &  0  &  0  &  0  &  \omega_b   \\
0 &  0  &  0  &  G_{mb}  & -\omega_b & -\gamma_b   \\
\end{pmatrix} ,
\end{equation}
where $\tilde{ \Delta}_m = \Delta_m + g_{mb} \langle q \rangle$ is the effective magnon-drive detuning including the frequency shift due to the magnomechanical interaction, and $G_{mb} = i \sqrt{2} g_{mb} \langle m \rangle$ is the effective magnomechanical coupling rate, where $\langle q \rangle = - \frac{g_{mb}}{\omega_b} |\langle m \rangle|^2 $, and $\langle m \rangle$ is given by
\begin{equation}\label{eq5}
\langle m \rangle =  \frac{  \Omega  ( i \Delta_a +\kappa_a) }{ g_{ma}^2 \! + ( i \tilde{ \Delta}_m + \kappa_m) ( i \Delta_a + \kappa_a) },
\end{equation}
which takes a simpler form 
\begin{equation}\label{avM}
\langle m \rangle \simeq  \frac{ i  \Omega  \Delta_a} {g_{ma}^2  - \tilde{ \Delta}_m \Delta_a }
\end{equation}
(a pure imaginary number) when $|\tilde{ \Delta}_m|, |\Delta_a| \gg  \kappa_a, \kappa_m$. The drift matrix in Eq.~\eqref{AAA} is provided under this condition. In fact, we will show later that $|\tilde{ \Delta}_m|, |\Delta_a| \simeq \omega_b  \gg  \kappa_a, \kappa_m$ [see Fig.~\ref{fig1} (b)] are optimal for the presence of all bipartite entanglements of the system. A similar finding has been observed in a hybrid atom-light-mirror system~\cite{DV08,JieAdP} due to the similarity of their Hamiltonians. Note that Eq.~\eqref{eq5} is intrinsically nonlinear since $\tilde{ \Delta}_m$ contains $ |\langle m \rangle|^2 $. However, for a given value of $\tilde{ \Delta}_m$ (one can always alter $\Delta_m$ by adjusting the bias magnetic field) $\langle m \rangle$, and thus $G_{mb}$, can be achieved straightforwardly.

Due to the linearized dynamics and the Gaussian nature of the quantum noises, the steady state of the quantum fluctuations of the system is a continuous variable (CV) three-mode Gaussian state, which is completely characterized by a $6\times6$ covariance matrix (CM) $\VV$ with its entries defined as $\VV_{ij}=\frac{1}{2}\langle u_i(t) u_j(t') + u_j(t') u_i(t)   \rangle$ ($i,j=1,2,...,6$). The steady-state CM $\VV$ can be achieved by solving the Lyapunov equation~\cite{DV07,Hahn}
\begin{equation}\label{Lyap}
A \VV+\VV A^T = -D,
\end{equation}
where $D={\rm diag} \big[ \kappa_a (2N_a+1), \kappa_a (2N_a+1), \kappa_m (2N_m+1),  \kappa_m (2N_m+1), 0,  \gamma_b (2N_b +1 ) \big]$ is the diffusion matrix, which is defined through $\langle  n_i(t) n_j(t') +n_j(t') n_i(t) \rangle/2 = D_{ij} \delta (t-t')$. To investigate bipartite and tripartite entanglement of the system, we adopt quantitative measures the logarithmic negativity $E_N$~\cite{LogNeg} and the residual contangle ${\cal R}_{\tau}$~\cite{Adesso}, respectively, where contangle is a CV analogue of tangle for discrete-variable tripartite entanglement~\cite{Wootters}. A {\it bona fide} quantification of tripartite entanglement is given by the {\it minimum} residual contangle~\cite{Adesso}
\begin{equation}
{\cal R}_{\tau}^{\rm min} \equiv {\rm min} \Big[ {\cal R}_{\tau}^{a|mb}, \, {\cal R}_{\tau}^{m|ab}, \,  {\cal R}_{\tau}^{b|am}  \Big],
\end{equation}
where ${\cal R}_{\tau}^{i|jk} \equiv C_{i|jk}-C_{i|j} -C_{i|k} \ge 0$ ($i,j,k=a,m,b$) is the residual contangle, with $C_{u|v}$ the contangle of subsystems of $u$ and $v$ ($v$ contains one or two modes), which is a proper entanglement monotone defined as the squared logarithmic negativity (see~\cite{SM} for more details of calculating $E_N$ and ${\cal R}_{\tau}$). A nonzero minimum residual contangle ${\cal R}_{\tau}^{\rm min}\,{>}\,0$ denotes the presence of {\it genuine} tripartite entanglement in the system.

\begin{figure}[t]
\hskip-0.18cm\includegraphics[width=\linewidth]{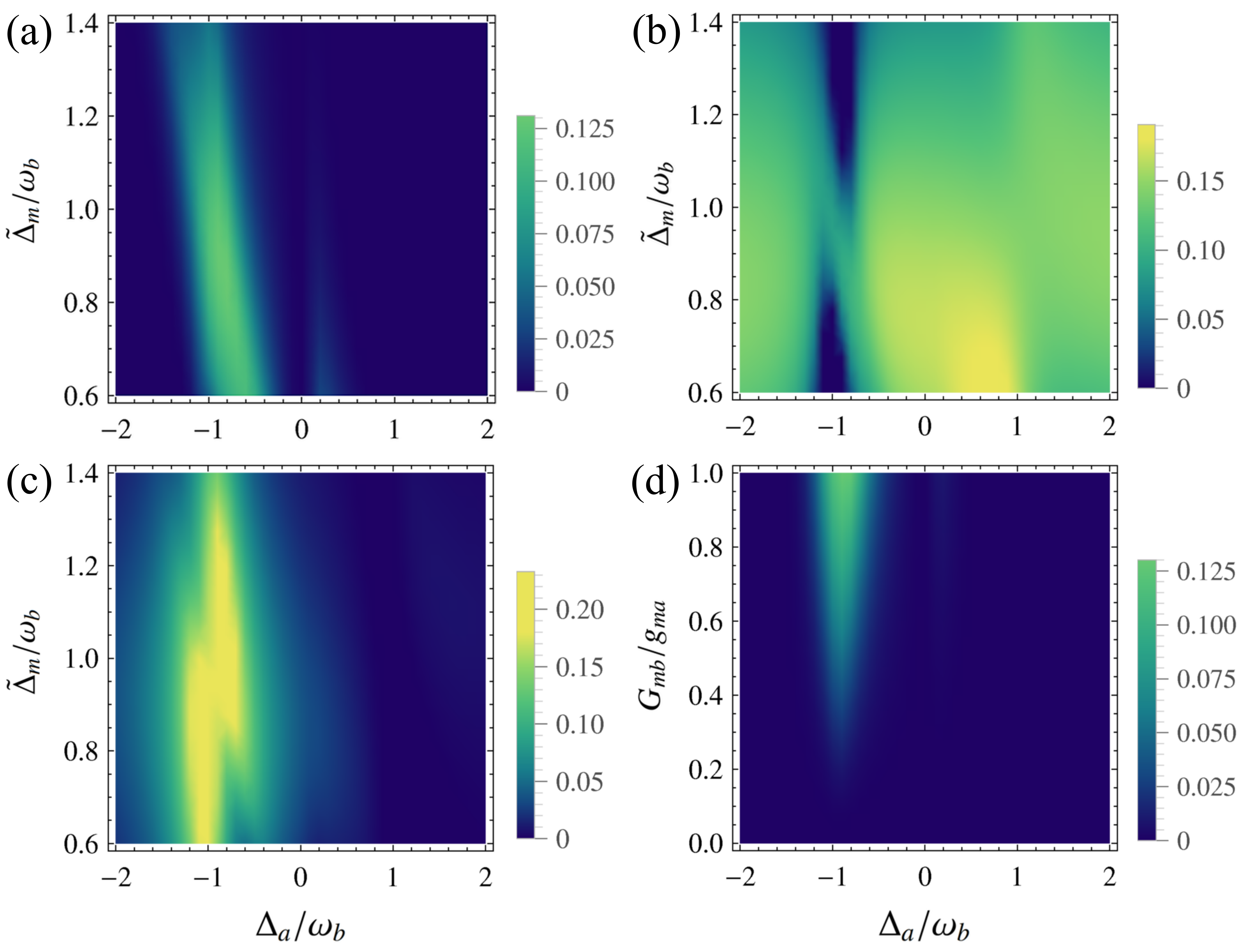} 
\caption{Density plot of bipartite entanglement (a) $E_{am}$, (b) $E_{mb}$, and (c) $E_{ab}$ versus detunings $\Delta_a$ and $\tilde{\Delta}_m$. (d) Density plot of $E_{am}$ versus $\Delta_a$ and the ratio of $G_{mb}/g_{ma}$ ($g_{ma}$ is fixed). The parameters are as in (a)-(c) but for $\tilde{ \Delta}_m=0.9\omega_b$. See text for the details of the other parameters.}
\label{fig2}
\end{figure}

The foremost task of studying entanglement properties in such a hybrid system is to find optimal detunings $\Delta_a$ and $\tilde{\Delta}_m$, i.e., to find optimal effective interactions among the three modes that can generate tripartite entanglement of them. In Fig.~\ref{fig2} (a)-(c), we show three bipartite entanglements versus detunings $\Delta_a$ and $\tilde{\Delta}_m$: $E_{am}$, $E_{mb}$, and $E_{ab}$ denote the cavity-magnon, magnon-phonon, and cavity-phonon entanglement, respectively. All results are in the steady state guaranteed by the negative eigenvalues (real parts) of the drift matrix $A$. It shows that there exists a parameter regime, around $\tilde{ \Delta}_m \simeq \omega_b$ and $\Delta_a \simeq -\omega_b$ [see Fig.~\ref{fig1} (b)], where all bipartite entanglements are present. In Fig.~\ref{fig2}, we have employed experimentally feasible parameters~\cite{Tang16}: $\omega_a/2\pi=10$ GHz, $\omega_b/2\pi=10$ MHz, $\gamma_b/2\pi=10^2$ Hz, $\kappa_a/2\pi=\kappa_m/2\pi=1$ MHz, $g_{ma}/2\pi =G_{mb}/2\pi =3.2$ MHz, and at low temperature $T\,{=}\,10$ mK. In this situation, $g_{ma}^2 \, {\ll}\, |\tilde{ \Delta}_m \Delta_a | \,\, {\simeq} \, \omega_b^2  $, the effective magnomechanical coupling $G_{mb} \, {\simeq} \sqrt{2} g_{mb} \frac{\Omega}{\omega_b}$ $\big[$see Eq.~\eqref{avM}$\big]$. $G_{mb}/2\pi =3.2$ MHz implies the drive magnetic field $B_0 \simeq 3.9 \times 10^{-5}$ T for $g_{mb}/2\pi \simeq 0.2$ Hz~\cite{Note}, corresponding to the drive power $P=8.9$ mW~\cite{SI}. 
In order to have all sizeable bipartite entanglements and at the same time keep the system stable, the two couplings $g_{ma}$ and $G_{mb}$ should be on the same order of magnitude and take moderate values. The physics of the optimal detuning $\tilde{ \Delta}_m \simeq \omega_b$ is as follows: The entanglement only survives with small thermal phonon occupancy. At this detuning, the magnomechanical (radiation pressure-like) interaction significantly cools the mechanical mode and simultaneously a considerable magnomechanical entanglement is generated due to the strong coupling~\cite{Note2}. The complementary distribution of the entanglement in Fig.~\ref{fig2} (b) and (a), (c) indicates that the initial magnon-phonon entanglement is partially transferred to the cavity-magnon and cavity-phonon subsystems, and this effect is prominent when the cavity detuning $\Delta_a \simeq -\omega_b$. Our hybrid system shows two advantages: (i) without involving the phonons the cavity photons and magnons interact via a beamsplitter interaction which yields zero entanglement between them. Nevertheless, by introducing the magnon-phonon interaction the cavity photons and magnons get entangled. This is clearly shown in Fig.~\ref{fig2} (d), where the cavity-magnon entanglement $E_{am}=0$ when $G_{mb}=0$ and $E_{am}$ increases with $G_{mb}$; (ii) thanks to the mediation of the magnons, the {\it indirectly} coupled cavity photons and phonons get entangled and the entanglement is even larger than those in directly coupled subsystems.

Note that the above results are valid only when the magnon excitation number $\langle m^{\dag} m \rangle \ll 2Ns=5N$. For a 250-$\mu$m-diameter YIG sphere, the number of spins $N \simeq 3.5 \times 10^{16}$, and $G_{mb}/2\pi =3.2$ MHz corresponds to $|\langle m \rangle| \simeq 1.1 \times 10^7$, and $\Omega \simeq 7.1 \times 10^{14}$ Hz, leading to $\langle m^{\dag} m \rangle \simeq 1.2 \times 10^{14} \ll 5N=1.8 \times 10^{17}$, which is well fulfilled. The strong magnon pump may cause unwanted nonlinear effects due to the Kerr nonlinear term $K m^{\dag}m m^{\dag} m$ in the Hamiltonian~\cite{You18,You16}, where $K$ is the Kerr coefficient, which is inversely proportional to the volume of the sphere. For a 1-mm-diameter YIG sphere used in Refs.~\cite{You18,You16}, $K/2\pi \approx 10^{-10}$ Hz~\cite{private}, and thus for what we use a 250-$\mu$m-diameter sphere, $K/2\pi \approx 6.4 \times 10^{-9}$ Hz. In order to keep the Kerr effect negligible, $K|\langle m \rangle|^3 \ll \Omega$ must hold. For the parameters used in Fig.~\ref{fig2}, we have $K|\langle m \rangle|^3 \simeq 5.7 \times 10^{13}$ Hz $\ll \Omega \simeq 7.1 \times 10^{14}$ Hz, implying that the nonlinear effects are negligible and the linearization treatment of the model is a good approximation.

\begin{figure}[t]
\hskip0.0cm\includegraphics[width=\linewidth]{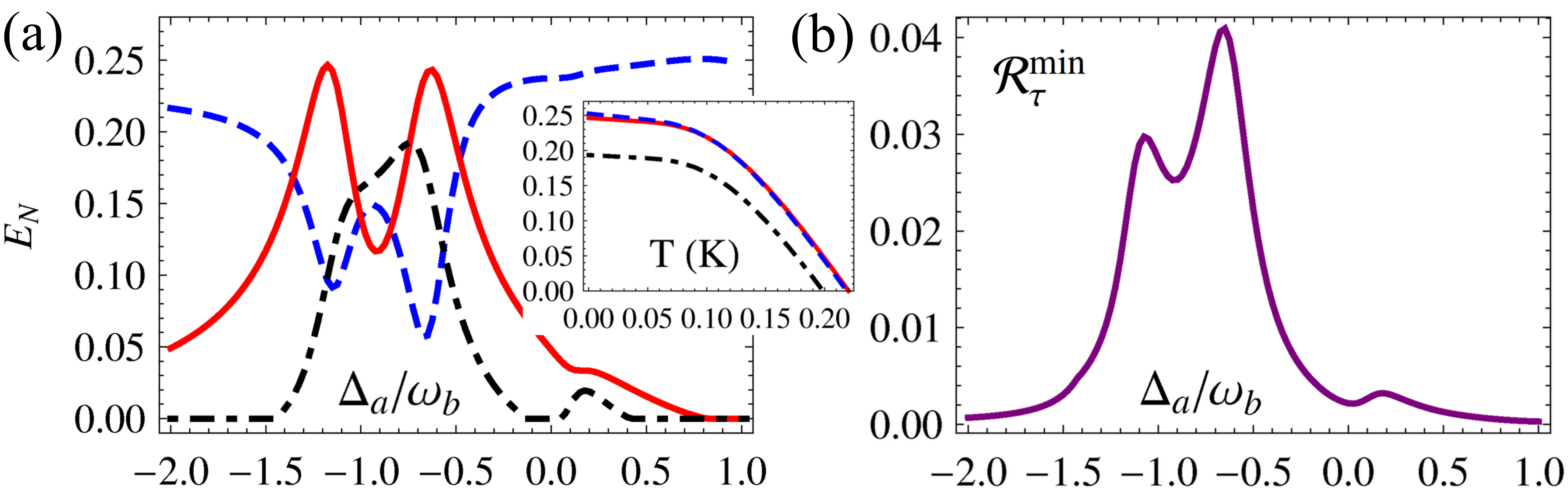}
\caption{(a) $E_{am}$ (dot-dashed), $E_{mb}$ (dashed), and $E_{ab}$ (solid) versus $\Delta_a$, and temperature (see the inset). In the inset $\Delta_a$ is optimized respectively for each bipartite entanglement. (b) Tripartite entanglement in terms of the minimum residual contangle ${\cal R}_{\tau}^{\rm min}$ versus $\Delta_a$. We take $G_{mb}/2\pi=4.8$ MHz and $\tilde{ \Delta}_m=0.9\omega_b$. The other parameters are as in Fig.~\ref{fig2}.}
\label{fig3}
\end{figure}

Figure~\ref{fig3} (a) shows more clearly the presence and interplay of the three bipartite entanglements. The parameters are as in Fig.2 but with a larger coupling rate $G_{mb}/2\pi=4.8$ MHz and an optimal detuning $\tilde{ \Delta}_m \simeq 0.9\omega_b$. 
All bipartite entanglements are robust against temperature and survive up to about $200$ mK, as shown in the inset of Fig.~\ref{fig3} (a). Apart from the simultaneous presence of all bipartite entanglements, the steady state of the system is also a {\it genuinely} tripartite entangled state, as demonstrated by the nonzero minimum residual contangle ${\cal R}_{\tau}^{\rm min}$ in Fig.~\ref{fig3} (b). Note that a 1.5 times larger $G_{mb}$ is used in Fig.~\ref{fig3} than in Fig.~\ref{fig2}, and hence $g_{mb}/2\pi \simeq 0.3$ Hz should be used to avoid the nonlinear effects with the same drive power.

\begin{figure}[t]
\hskip0.0cm\includegraphics[width=\linewidth]{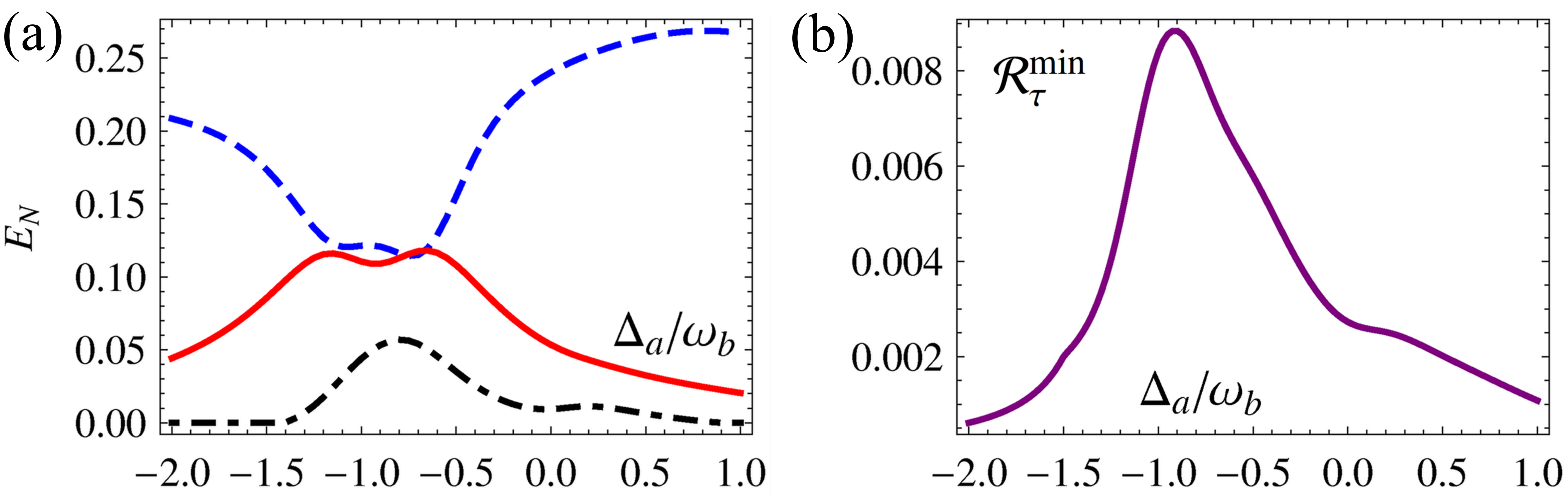}
\caption{(a) $E_{am}$ (dot-dashed), $E_{mb}$ (dashed), and $E_{ab}$ (solid) versus $\Delta_a$. (b) Minimum residual contangle ${\cal R}_{\tau}^{\rm min}$ versus $\Delta_a$. The parameters are as in Fig.~\ref{fig3} except for $\kappa_a/2\pi=3$ MHz and $\kappa_m = \kappa_a/5$. }
\label{fig4}
\end{figure}

Lastly, we discuss how to detect and verify the entanglement. The generated tripartite or bipartite entanglement can be verified by measuring the corresponding CMs~\cite{enOM,DV07}. The cavity field quadratures can be measured directly by homodyning the cavity output. The magnon state can be read out by sending a weak microwave probe field and by homodyning the cavity output of the probe field. This requires that the dissipation rate of the magnon mode should be much smaller than that of the cavity mode, such that when the drive is switched off and all cavity photons decay the magnon state remains almost unchanged, at which time a probe filed is sent. Figure~\ref{fig4} shows the entanglements for the case of $\kappa_a=5\kappa_m$, where tripartite entanglement can still be achieved. Finally, the mechanical quadratures can be measured by coupling the YIG sphere to an additional optical cavity which is driven by a weak red-detuned  light. In this situation, the optomechanical interaction is effectly a beamsplitter interaction which maps the phonon state onto the cavity output field~\cite{Jie18}.

In conclusion, we have presented a scheme to generate tripartite entanglement in a cavity magnomechanical system where a microwave cavity mode is coupled to a magnon mode in a YIG sphere, and the latter is simultaneously coupled to a mechanical mode via magnetostrictive force. We have shown that with experimentally reachable parameters cavity photons, magnons, and phonons can be entangled with each other and the steady state of the system exhibits genuine tripartite entanglement. We have also provided possible strategies to measure the entanglement. Our scheme will open new perspectives for the realization of quantum interfaces among microwave, magnonic, and mechanical systems serving for the quantum information processing, where the mechanical oscillator can act as storage of information which can be transferred to other systems leading to hybridization.  Our work suggests the possibility of several lines of investigation, for example the study of tripartite entanglement in magnon-photon-superconducting qubit systems~\cite{Naka17} where the quantized states of magnons have been observed. It should be possible to prepare the magnon system in a variety of nonclassical states by suitably driving it or by using nonlinear collective interaction~\cite{GA97,10qubit} quadratic in spin operators.



{\it Acknowledgments}. We thank J. Q. You, C.-L. Zou, and D. Vitali for helpful discussions. This work has been supported by the National Key Research and Development Program of China (Grants No. 2017YFA0304202 and No. 2017YFA0304200) and the Biophotonics program of the Texas A{\rm \&}M University.

\section*{SUPPLEMENTARY MATERIAL}

\section*{I. Derivation of the Rabi frequency of the magnon drive}

The Rabi frequency $\Omega$ denotes the coupling strength of the drive magnetic field with the magnon mode. We now derive its specific expression as follows. This is important because it will be used later on to determine the effective magnomechanical coupling rate and examine the validity of our linearized model. 

The Hamiltonian for a spin in a magnetic field is $H = - \gamma \vec{s} \cdot \vec{B}$, where $\vec{s}$ is the spin angular momentum. Since the YIG sphere contains a large number of spins, we define the collective spin angular momentum $\vec{S}=\sum \vec{s}$. Therefore, the Hamiltonian for the spins in the drive magnetic field (e.g., along the $y$ direction) with amplitude $B_0$ and frequency $\omega_0$ is given by $H_d = - \gamma \vec{S} \cdot \vec{B} = - \gamma S_y B_0 \cos \omega_0 t$, where $\vec{S} =(S_x, S_y, S_z)$. $H_d$ can be written in terms of the raising and lowering operators $S^{\pm}$, $S^{\pm} = S_x \pm i S_y$, i.e., $H_d = i \frac{\gamma B_0}{4} (S^+ -S^-) (e^{i \omega_0 t} + e^{-i \omega_0 t})$. The collective spin operators $S^{\pm}$ are related to the bosonic annihilation and creation operators of the magnon mode, $m$ and $m^{\dag}$, via the Holstein-Primakoff transformation, $S^+ = \hbar \sqrt{2Ns - m^{\dag} m}\, m$ and $S^- = \hbar \, m^{\dag}\! \sqrt{2Ns - m^{\dag} m}$~\cite{HPT}, where $N$ is the total number of spins and $s=\frac{5}{2}$ is the spin number of the ground state Fe$^{3+}$ ion in YIG. For the low-lying excitations, $\langle m^{\dag} m \rangle \ll 2Ns$, the above transformations can be approximated as $S^+ \approx \hbar \sqrt{5N}\, m$ and $S^- \approx \hbar \sqrt{5N}\, m^{\dag} $. This leads to the Hamiltonian 
\begin{equation}
\begin{split}
H_d/\hbar &= i \frac{\sqrt{5}}{4} \gamma \! \sqrt{N} B_0 \, (m -m^{\dag}) (e^{i \omega_0 t} + e^{-i \omega_0 t})  \\
&\approx  i \Omega \, (m e^{i \omega_0 t}  -m^{\dag}  e^{-i \omega_0 t}), 
\end{split}
\end{equation}
where $\Omega =\frac{\sqrt{5}}{4} \gamma \! \sqrt{N} B_0$, $\gamma/2\pi= 28$ GHz/T, $N=\rho V$ with $V$ the volume of the sphere and $\rho=4.22 \times 10^{27}$ m$^{-3}$ the spin density of the YIG, and for taking ``$\approx$" we have made the rotating-wave approximation.

\section*{II. Quantification of Gaussian bipartite and tripartite entanglement}

We adopt the logarithmic negativity~\cite{LogNeg} for quantifying bipartite entanglement of our three-mode Gaussian state, which is defined as
\begin{equation}\label{LOGNEG}
E_N \equiv \max[0, \, -\ln2\tilde\nu_-],
\end{equation}
where $\tilde\nu_-=\min{\rm eig}|i\Omega_2\tilde{\VV}_4|$ (with the symplectic matrix $\Omega_2\,{=}\oplus^2_{j=1} \! i\sigma_y$ and $\sigma_y$ is the $y$-Pauli matrix) is the minimum symplectic eigenvalue of the CM $\tilde{\VV}_4={\cal P}_{1|2}{\VV_4}{\cal P}_{1|2}$, where $\VV_4$ is the $4\times 4$ CM of two subsystems, obtained by removing in $\VV$ the rows and columns of the uninteresting mode, and ${\cal P}_{1|2}={\rm diag}(1,-1,1,1)$ is the matrix that realizes partial transposition at the level of CMs~\cite{Simon}.

For the study of tripartite entanglement, we adopt a quantitative measure the residual contangle ${\cal R}_{\tau}$~\cite{Adesso}, given by
\begin{equation}
{\cal R}_{\tau}^{i|jk} \equiv C_{i|jk}-C_{i|j} -C_{i|k}, \,\, (i,j,k=a,m,b) 
\end{equation}
where $C_{u|v}$ is the contangle of subsystems of $u$ and $v$ ($v$ contains one or two modes), which is a proper entanglement monotone defined as the {\it squared} logarithmic negativity~\cite{Adesso}. For calculating the {\it one-mode-vs-two-modes} logarithmic negativity $E_{i|jk}$, one only needs to follow the definition of Eq.~\eqref{LOGNEG} simply by replacing $\Omega_2\,{=}\oplus^2_{j=1} \! i\sigma_y$ with $\Omega_3\,{=}\oplus^3_{j=1} \! i\sigma_y$, and $\tilde{\VV}_4={\cal P}_{1|2}{\VV_4}{\cal P}_{1|2}$ with $\tilde{\VV}={\cal P}_{i|jk}{\VV}\,{\cal P}_{i|jk}$, where ${\cal P}_{1|23}={\rm diag}(1,-1,1,1,1,1)$, ${\cal P}_{2|13}={\rm diag}(1,1,1,-1,1,1)$, and ${\cal P}_{3|12}={\rm diag}(1,1,1,1,1,-1)$ are partial transposition matrices. The residual contangle satisfies the monogamy of quantum entanglement, ${\cal R}_{\tau}^{i|jk} \ge 0$, i.e.,
\begin{equation}
C_{i|jk} \ge C_{i|j} +C_{i|k},
\end{equation}
which is similar to the Coffman-Kundu-Wootters monogamy inequality~\cite{Wootters} hold for the system of three qubits.

A {\it bona fide} quantification of CV tripartite entanglement is provided by the {\it minimum} residual contangle~\cite{Adesso}
\begin{equation}
{\cal R}_{\tau}^{\rm min} \equiv {\rm min} \Big[ {\cal R}_{\tau}^{a|mb}, \, {\cal R}_{\tau}^{m|ab}, \,  {\cal R}_{\tau}^{b|am}  \Big],
\end{equation}
which ensures that ${\cal R}_{\tau}^{\rm min} $ is invariant under all permutations of the modes and is thus a genuine three-way property of any three-mode Gaussian state.


\begin{thebibliography}{99}
\bibitem{Note}
We consider a larger value of  $g_{mb}$ than that measured in the experiment of Ref.~\cite{Tang16} in order to lower the pump power to avoid unwanted nonlinear effects. The coupling strength can be increased by using smaller YIG spheres.
\bibitem{SI}
Note that time average of energy per unit volume is $\frac{B_0^2}{2\mu_0}$ ($\mu_0$ is vacuum magnetic permeability) and hence power $P=\frac{B_0^2}{2\mu_0} Ac$, where $c$ is the speed of an electromagnetic wave propagating through the vacuum and $A$ is the cross-sectional area, for which we take the maximum value $\pi R^2$, with $R$ being the radius of the YIG sphere. Therefore, $B_0=\frac{1}{R} \sqrt{\frac{2P\mu_0}{\pi c}}$.
\end{thebibliography}
\end{document}